\documentclass[12pt]{article}
\usepackage{amsmath,amssymb}
\usepackage[dvips]{graphicx}
\usepackage{pdfpages}
\usepackage{pdflscape}
\usepackage{lmodern}
\usepackage{extarrows}
\usepackage{rotating}
\usepackage{MnSymbol}
\usepackage{caption}
\usepackage{float}

\begin{document}

\begin{center}

\section*{The study of the energy spectrum\\ of a system of quantum micro-vortices \\in a bounded spatial domain.}

{S.V. TALALOV}

{Department of Applied Mathematics, Togliatti State University, \\ 14 Belorusskaya str.,
 Tolyatti, Samara region, 445020 Russia.\\
svt\_19@mail.ru}

\end{center}

\begin{abstract}
This study focuses on microscopic-sized quantum vortex filaments that are shaped like a circle. The model we considered examines loops with different radii and a small but non-zero core diameter. These loops are located in a bounded domain $D$. The quantization scheme of the  classical vortices is based on the new  approach proposed by the author \cite{Tal22_1,Tal24_2}.
For these loops, we calculate both the quantized circulation and the energy spectrum, which are  perfectly non-trivial. 
To understand how the results we have obtained can be used to describe the initial stage of turbulence in a quantum fluid, we study  a system of $K$ random, non-interacting vortices.
 We explain how specific energy and circulation spectra lead to the occurrence of turbulence in the context of the developed approach.
\end {abstract}

\vspace{5mm}

{\bf keywords:}   quantum vortices,  circulation quantization, quantum turbulence.
 
\vspace{5mm}

 {\bf PACS numbers:}   47.10.Df    ~~47.32.C

\vspace{5mm}

\section{Introduction}

\paragraph~

~~~ The general direction of the research in this paper can be related to quantum hydrodynamics.
More specifically, in this study, we examine the energy spectrum of quantum vortex loops and investigate the potential causes of turbulence.
The approach proposed by the author to vortex quantization lies outside well-known models
(for example, Gross-Pitaevskii model). 
The developed theory also doesn't aim to improve our understanding of the superfluidity phenomenon. As R. Feynman pointed out in \cite{Feyn_1}, quantum hydrodynamics does not explain superfluidity.   
The author hopes that the proposed theory will be a certain advance in the study of quantum turbulence at initial stage. Indeed,  it was established many years ago,  that vortices play a crucial role in understanding the nature of this phenomenon \cite{Feyn_2}
(see also \cite{Vinen,Tsu_1,Sonin,Baren}).
The theory of quantum vortices has a long and fascinating history. However, this paper  is not intended to be a comprehensive review of this topic.
	Here we note only some  fundamental books of \cite{Sonin,Nemir,Donn}  that reflect different periods of research and approaches to the problem.
Among the recent studies, we can highlight the works \cite{Sb1, Sb2}, where quantum vortices were described using a modified Navier-Stokes equation.
We also want to mention an article  \cite{GladW}  in which a numerical analysis of the interaction of vortices is conducted using the Gross-Pitaevsky theory.

The initial stage of turbulence can be described as the formation of individual micro-vortices that do not interact with each other.
In this paper, we will not focus on the process of creating vortices ''out of nothing''. These issues have been discussed in other works by the author, which we will come later.
We will also not cover all the issues that arise when trying to describe a turbulent flow.
In general, despite the extensive work being done on this issue, it still remains a problem \cite{PoMuKr_1}.  
 Instead, we will focus on a system of microscopic and mesoscopic  vortex rings, which can be seen as the first stage of turbulent flow formation.

Our method for developing a quantum theory of a vortex ring follows the standard approach in quantum theory. The new and unexpected findings are due to two factors: the unusual choice of quantization variables\footnote{It's worth noting that choosing non-standard variables for quantization can lead to unexpected outcomes, even for simple systems like a harmonic oscillator \cite{Kast}.} and the unusual way of calculating the energy of a vortex ring. 
On a classical level, we start from Local Induction Approximation for thin vortex filaments which are described by the function $\boldsymbol{r}(t , S)$. This   function  means the the filament coordinates where symbol $S$ denotes  the natural parameter. 
The well-known equation
\begin{equation}
        \label{LIE_eq}
       { \partial_{t^\ast} {\boldsymbol{r}}({t^\ast} , S)} ~=~ \beta_1\,
       { \partial_S{\boldsymbol{r}}({t^\ast} ,S)}\times {\partial_S^{\,2}{\boldsymbol{r}}({t^\ast},S)}\,
        \end{equation}
is deduced if we replace the
  physical time $t$ with the ''evolution parameter''   $t^\ast$ which is defined as $ t^\ast =  {t\Gamma  }/{4\pi}\,.$	
				Symbol $\Gamma$ means the circulation.

		The equation (\ref{LIE_eq}) can be derived using a specific regularization technique with parameter $\ell$. As a consequence, the parameter $\beta_1 = \ln({2\ell}/{\sf a}) - 1$  appears in Eq.(\ref{LIE_eq}). The value ${\sf a}$ is small but non-zero radius of the filament core.  
			We will not go into detail here, as you can find more information in Book \cite{AlKuOk}, for example.  		For our  subsequent studies, we assume  $\beta_1 = const$. 
To take into account the flow inside the core, we need to add more terms to Eq.(\ref{LIE_eq}). We'll talk about them later.

		When creating a quantum theory of micro-vortices, it is crucial to understand the concept of vortex ring energy. 	So,  the standard formula for the energy \cite{Saffm} leads to the ambiguous result for the filaments with a small core radius  ${\sf a}$. This is especially important when we try to take into account  the energy contribution of a non-zero flux inside the core. 		
	Indeed  {\it''\dots  considering how small the vortex core in helium II is, i.e., of order an angstrom, it would seem that one either ought to know how it is constructed or one ought to find a way to ignore it. Unfortunately neither goal has been achieved''} \cite{Donn}.

	We apply a group-theoretic approach to define  the full energy of a thin  (${\sf a} \to 0$, ${\sf a} \not= 0$ ) vortex filament. This approach was first introduced by the author in his earlier work
	\cite{Tal22_1}, which explored the small oscillations of a vortex ring at the quantum level in an infinite space. In his later works, the author developed the theory, including for some special domains. You can find more information about this in the paper \cite{Tal24_2}, for example.
	Among other things, the proposed approach allows us to naturally use the quantum theory of many bodies to describe how vortices interact. We will not go into any detail about the interaction between vortices here.

		In order to implement this approach, we extend the space-time symmetry group of the theory to the central-extended Galilei group $\widetilde{\mathcal G}_3$. Lee algebra of the  group $\widetilde{\mathcal G}_3$ has three Cazimir functions:  
		 \[ {\hat C}_1 = \mu_0 {\hat I}\,,\quad 
  {\hat C}_2 = \left({\hat M}_i  - \sum_{k,j=x,y,z}\epsilon_{ijk}{\hat P}_j {\hat B}_k\right)^{\!2} 
  \quad {\hat C}_3 = \hat H -  \frac{1}{2\mu_0}\sum_{i=x,y,z}{\hat P}_i^{\,2}\,,\]                        
       where        ${\hat I}$ is the unit operator,     ${\hat M}_i$,   $\hat H$,  ${\hat P}_i$         and  ${\hat B}_i$  ($i = x,y,z$)
        are the respective generators of rotations, time and space translations and Galilean boosts, value $\mu_0$ is a central charge. 
		Traditionally, the function  ${\hat C}_3 $  can be interpreted as  an  ''internal energy of the particle'' as well as the central charge $\mu_0$ is usually interpreted as  ''mass''.  This allows us to define energy in classical theory using the formula
		\begin{equation}
		\label{E_general}
 {\cal E}_{cl} ~=~ \frac{{\bf p}^{\,2}}{2\mu_0} ~+~ {\hat C}_3(\varpi, \chi, \dots)\,,
\end{equation}
where ${\bf p}$ is vortex monentum an $\varpi, \chi, \dots$ some set of the ''internal'' variables.
	In the next section, we will apply the principles we have discussed to describe a classic dynamic system called here  as a "micro-vortex".

\section{ The classical system we are discussing}

At the first stage, we  investigate circular vortex ring of an arbitrary  radius $R$.
We believe that the value $R$  is quite small since our ultimate goal is to build a quantum theory of microscopic and mesoscopic closed vortices.
Nevertheless, we will consider the restriction on the radius of the ring to be in its most general form: we assume that inequality
\begin{equation}
        \label{R_f}
 R_f ~\le~ R ~<~ R_{max} 
 \end{equation}
holds for certain values $R_f$ and $R_{max}$. The constant $R_f$  depends on the type of fluid we are considering.
For example,  the constant  $R_f$ may be related to the size of a fluid molecule, inter-atomic distance
or  size of a stable molecular cluster. This constant is a fundamental constant in our theory.
The reason for the existence of the constant $R_{max}$  is because we are considering vortices in a bounded  domain. Furthermore, if the radius is sufficiently large, the specific quantum properties of the vortex can be destroyed due to thermal fluctuations.
We consider that the inequality  $R < R_{max}$  holds, but we don't think of this constant as fundamental in this context.  
Other fundamental constants are fluid density $\varrho_0$ and the speed of sound in a fluid $v_0$.
We  also use the constant $L$, which describes the size of the domain  $D$.
In certain cases, the ratio  $R_{max} \sim L/2$ is fulfilled.  We will discuss this issue later.
To simplify the formulas, we introduce the notations
\[  t_0 = \frac{L}{v_0}\,, \qquad    {\mathcal E}_0 = \mu_0 v_0^2\,,\]
These  constants  define the  time and energy scales in constructed classical  theory. 
  The mass parameter $\mu_0$ is a central charge for the central extended Galilei group.  The relevance  of this group for our model  was discussed in the cited author's works in detail.  We  use the variable parameter $\mu_0$ together with  ''natural'' mass parameter $\tilde\mu_0 = \pi\rho_0 R_f^3$.

Exploring the dynamics of the vortex ring, it will be more convenient  to use dimensionless parameters $\tau$ and $\xi$ instead of the evolution parameter   $ t^\ast =  {t\Gamma  }/{4\pi}$ (where symbol 
   $t$ means ''real time'', see Eq.(\ref{LIE_eq})) and the natural parameter $S$:

\begin{equation}
        \label{tau_s}
 \tau    ~=~ \frac{t^*}{R^2} ~\equiv~  \frac{t\Gamma  }{4\pi R^2}\,,	\qquad \quad \xi ~=~ \frac{S}{R}\,, \qquad
\xi \in [0,2\pi]\,.
 \end{equation}
Thus, we  consider the following equation for a projective vector 
${\mathfrak r} =  {\boldsymbol{r}}/R$:

\begin{eqnarray}
        \label{LIE_pert}
        \partial_\tau {\mathfrak r}(\tau ,\xi)  & = &
        \beta_1 \Bigl(\partial_\xi{\mathfrak r}(\tau ,\xi)\times\partial_\xi^{\,2}{\mathfrak r}(\tau ,\xi)\Bigr)   ~+ \nonumber \\
				~~ & + & \omega\Bigl(2\,\partial_\xi^{\,3}{\mathfrak r}(\tau ,\xi) ~+~ 
        {3}\,\bigl\vert\, \partial_\xi^{\,2}{\mathfrak r}(\tau ,\xi)\bigr\vert^{\,2}\partial_\xi{\mathfrak r}(\tau ,\xi)\Bigr)\,,
				        \end{eqnarray}
where the values $\beta$   and  $\omega$ are finite dimensionless constants.
{ In total,  parameter $\omega$  is determined by the velocity of fluid flow within the vortex core.}
A detailed deducing  of this equation { as well as deducing the explicit expression of the parameter $\omega$  from the conventional physical values} was made in  the book \cite{AlKuOk}.

In this paper,  we use the representation of any smooth closed curve of length $S_0$   in the form
\begin{equation}
        \label{proj_r}
                                   {\mathfrak r}(\tau,\xi) ~=~  \frac{\boldsymbol{q}}{R} ~+~ 
          \int\limits_{0}^{2\pi}  \left[\, {\xi - \eta}\,\right] {\boldsymbol j}(\tau,\eta) d\eta\,, \qquad  R = \frac{S_0}{2\pi} \,,
                  \end{equation}
where vector ${\boldsymbol{q}} = {\boldsymbol{q}}(\tau)$ defines the position of the curve in the coordinate system (the   center of the circular vortex ring, for example).
The notation $[\,x\,]$ means the integer part of the number $x/{2\pi}$:
\begin{equation}
        \label{int_part}
 [\,0\,] = 0\,, \qquad  [\,x+2\pi\,] = [\,x\,]+1\,, \qquad \forall\,x\,.
\end{equation}
 Vector ${\boldsymbol j}(\tau,\eta)$ is the unit  tangent (affine) vector for this curve. Indeed, 
${\partial {\boldsymbol{r}}}/{\partial S} ~=~  {\boldsymbol j}$
in accordance with  formulas   (\ref{tau_s}),    (\ref{proj_r}) and   (\ref{int_part}).
The  components  of the vector-function ${\boldsymbol j}(\tau,\xi)$  satisfy  to the conditions

            \begin{equation}
  \label{constr_j_0}
    \int\limits_{0}^{2\pi}{\boldsymbol j}(\xi)\,d\xi  ~=~ 0\,.
   \end{equation}      
These conditions  ensure the closeness  of the curve in question: indeed,  the equality $ {\mathfrak r}(\tau,\xi + 2\pi) = {\mathfrak r}(\tau,\xi) $ will be fulfilled.
Since we are interested in solutions of the  equation (\ref{LIE_pert}) in the form of closed curves, we will use representation (\ref{proj_r}) for such solutions.

The  ''master'' equation   (\ref{LIE_pert}) has the exact solution

\begin{equation}
        \label{our_sol}
 {\mathfrak r}(\tau ,\xi) ~=~ \Bigl(\, \frac{q_x}{R} +  \cos(\xi +\phi)\,,\quad \frac{q_y}{R} +  \sin(\xi +\phi  )\,, \quad \frac{q_z}{R} + \beta \tau \,\Bigr)\,, 
\end{equation}
where $\phi \equiv \phi(\tau) =  \phi_0 +  \omega\tau$.  Later in this paper, we will be looking into this  simplest  particular solution.
{
In our opinion, it is suitable for modeling closed micro-vortices. Let's make a comment on other possible solutions for Eq.(\ref{LIE_pert}).
It is known that this equation   is equivalent to Hirota's equation (see 
 book \cite{AlKuOk} for example).  The special case of Eq.(\ref{LIE_pert}) where the constant $\omega=0$, corresponds to the nonlinear Schrödinger equation \cite{Hasim}.  These non-linear integrable equations have soliton solutions, which can be found using the inverse scattering transform method.  As opposed to case of an infinite vortex filaments, when $\xi \nolinebreak \in \nolinebreak (-\infty,\infty)$,   the study of periodic solutions of such equations  needs complicated mathematical methods \cite{Dubrovin}.     We also face significant additional challenges here in achieving our final goal: creating quantum models based on these solutions. These issues are beyond the scope of this article. The author hope to explore this problem in the future.  Of course, we can also consider the small oscillations of the circular ring. This was done in the author's previous research on this topic. However, in this paper, we don't do that to simplify the final formulas. }

Let us define the value $\Delta R$:
\begin{equation}
        \label{Delta_R}
	\Delta R ~=~ \sqrt{R^2 - R_f^2}\,,
				\end{equation}
			where the constant $R_f$ was introduced by the inequality (\ref{R_f}).	

To describe the excitations of the vortex micro-ring we are considering, we add the dynamical variables: the value $\Delta R$, phase $\phi$, the coordinates ${\boldsymbol q}$  and the components of the unit vector ${\boldsymbol{e}_z}$. 
Because the vortex in question is oriented in space in any direction, the vector ${\boldsymbol{e}_z}$ has two independent components. 
 Also, we consider the circulation $\Gamma$ as an additional dynamic variable. This step, which is characteristic of the proposed approach, allows us to account for the movement of the surrounding fluid in a minimal way.  From the perspective of developing a more advanced theory, this step will enable us to define a wider range of variables that describe the dynamics of vortices.

In addition to the Eq. (\ref{LIE_pert})  that describes the dynamics of a vortex filament, we also assume that the well-known formula \cite{Batche}  for the canonical momentum ${\boldsymbol p}$ is fulfilled:
	  
   \begin{equation}
        \label{p_and_m_st}
        {\boldsymbol p} ~=~ \frac{\varrho_0}{2 }\,\int\,\boldsymbol{r}\times\boldsymbol{w}(\boldsymbol{r})\,dV\,.
        \end{equation}

The vector  ${\boldsymbol{w}}(\boldsymbol{r})$  means  the vorticity and the constant 
$\varrho_0$ means the  fluid density.                
      As usual,   the vorticity of the  closed vortex filament is calculated as follows
   \begin{equation}
        \label{vort_w}
     {\boldsymbol{w}}(\boldsymbol{r}) ~=~  \Gamma
                  \int\limits_{0}^{2\pi}\,\hat\delta(\boldsymbol{r} - \boldsymbol{r}(\xi))\partial_\xi{\boldsymbol{r}}(\xi)d\xi\,,
       \end{equation}            
   where the symbol $\Gamma$ denotes the circulation  and the symbol  $\hat\delta(\xi)$ means $2\pi$-periodical $3D$  $\delta$-function.
We are only considering circular vortex rings with a radius of $R$ (see Eq.(\ref{our_sol})), so this formula applies:
\begin{equation}
        \label{p_Gamma}
	{\boldsymbol{p}}  ~=~    \pi\varrho_0 {R}^2 \Gamma        {\boldsymbol e}_z 
	\end{equation}

To complete the  construction of  the set of the independent variables, we define the dimensionless variables $\varpi$ and
 $\chi$ instead of variables $\Delta R$ and $\phi(\tau)$. 
The relevant definitions are as follows:

\[ \chi   ~=~ \frac{\Delta R}{R_f}\cos(\phi_{\,0} +\omega\tau)\,, \qquad \varpi  ~=~ \frac{\Delta R}{R_f}\sin(\phi_{\,0} +\omega\tau)\,.\]
Clearly, the behavior of these variables is similar to that of a harmonic oscillator.
Finally, we  postulate the set
\begin{equation}
\label{new_set1}
{\cal A}^{\,\prime} ~=~ \bigl\{\, {\boldsymbol p}\,,  {\boldsymbol{q}}\,;~ \varpi\,, \chi\,
  \,\bigr\}
 \end{equation}
as the set of fundamental independent variables for the considered theory.
As mentioned above, in this article we do not consider small  oscillations (Kelvin waves) of the vortex ring. We are only interested in rings that are circular in shape, with different radii.  In their early works \cite{Tal22_1, Tal24_2}, the authors explored the small oscillations of ring-shaped vortices using the proposed approach.

Thus, 
 the circulation $\Gamma$ becomes  the  function  of the   variables (\ref{new_set1}). This function is defined from the equation
\begin{equation}
\label{main_con}
{\boldsymbol p}^{\,2}   ~=~  \pi^2\varrho_0^2 {R_f}^{\,4}\Bigl(1 + \varpi^{\,2} + \chi^{\,2} \Bigr)^{2}\, \Gamma^{\,2}\,.
\end{equation}

It is important to note that the dynamical system we are discussing is simplified to a "particle + oscillator" system only when it evolves in ''conditional time'' $t^\# =  \tau t_0$. 
The author's previous works have carefully developed and researched the Hamiltonian structure of the theory that supports this claim.  We will not go into detail about it here.  In the final formulas, we will have to return to real time $t$.


\section{Quantum microvortices}

Taking into account the formula (\ref{new_set1}), it is natural to assume that  the quantum states of the considered vortex loop are the vectors of the Hilbert space
\begin{equation}
	\label{space_quant}
	\boldsymbol{H}_1  ~=~  \boldsymbol{H}_{pq} \otimes   \boldsymbol{H}_b  \,,
	\end{equation}
			where the symbol   $\boldsymbol{H}_{pq}$  denotes the Hilbert space  of a free structureless particle in the domain  $D \subset {\sf R}_3$.  
	{	We suppose that the boundary $\partial D$  of this domain is a certain piecewise-smooth surface in the space 	${\sf R}_3$.}  			
			In this paper we assume
			$\boldsymbol{H}_{pq} = L_2(D)$.  	The	symbol $\boldsymbol{H}_b $ denotes		the Hilbert space  of the quantized harmonic oscillator with classical variables $\chi$ and $\varpi$ (see definitions of these quantities before the formula (\ref{new_set1})).
				This space is formed by the vectors
		\[|\,n\rangle   ~=~  \frac{1}{\sqrt{n!}} (\hat{b}^+)^n    |\,0_b\rangle   \qquad 
		[\,\hat{b}, \hat{b}^+] ~=~ \hat{I}_b\,, \quad \hat{b}|\,0_b\rangle ~=~ 0\,,  \]
			where vector  $|\,0_b\rangle \in \boldsymbol{H}_b$ is vacuum vector and symbols $\hat{b}^+$ and $\hat{b}$ mean the creation and annihilation operators. 	
On this ground, we quantize the dimensionless variables  $\varpi$ and $\chi$   as follows:
\[ \chi ~\to~ \sigma_{ph}\, \frac{\hat{b} + \hat{b}^+}{\sqrt{2}}\,, \qquad 
\varpi ~\to~ \sigma_{ph}\, \frac{\hat{b} - \hat{b}^+}{i\sqrt{2}}\,,  \]
where the dimensionless constant
\[\sigma_{ph} ~=~ \sqrt{\frac{{\sf h}}{{\mathcal E}_0 t_0}} ~=~ \sqrt{\frac{{\sf h}}{\mu_0 v_0 L}}\,\]
depends on the Planck constant ${\sf h}$. The value ${\mathcal E}_0 t_0$ defines the scale of action in our classical theory.
The variables $\boldsymbol q $ and $\boldsymbol p $ are quantized according to the standard rules of quantum mechanics for a free non-relativistic particle in  Hilbert space $L_2(D)$.

The first non-trivial result is that the radius of the vortex ring is quantized.
Indeed, considering the definition of the variables $\chi$ and $\varpi$, as well as Eq.(\ref{Delta_R}), we can derive the following expression for the quantized vortex radius $R$: 
\[ R^{\,2} ~\longrightarrow~ \hat{R}^{\,2} ~=~ R_f^{\,2}
\left[{\hat I}_b + \sigma_{ph}^2\left({\hat b}^+ {\hat b} + \frac{1}{2}\right)\right]\,.\]   
Therefore, the operator $\hat{R}$, which is a well-defined operator in the space $\boldsymbol{H}_b$, has the following eigenvalues  ${R}_n$:
\begin{equation}
\label{spectrum_R}
 {R}_n ~=~ R_f\sqrt{1 + \sigma_{ph}^2\left(n + \frac{1}{2}\right)} \,,
\qquad  n ~=~ 0,1,\dots, N\,.
\end{equation}
The finite number $N$ appears here because the values ${R}_n$ were restricted by the inequalities (\ref{R_f}).   In light of the discussion about the value of $R_{max}$ at the beginning of section 1, it seems interesting to examine how the spectrum behaves when the number $n$ is large. 
	The result comes from  simple transformations:
		\[ R_{n+1} - R_n ~\simeq~ \frac{\sigma_{ph} R_f}{2\sqrt{n}}\,.\] 
	Therefore, this spectrum becomes quasi-continuous for large $n$.	
	Later, we will discuss the number $N$.

	Next, we return to the equation (\ref{main_con}).
	This equality can be seen as an implicitly defined function $\Gamma = \Gamma({\boldsymbol p};\varpi,\chi)$. 		
		According to our established rules for quantization, we have  the following spectral problem for possible values of circulation $\Gamma$:
\begin{equation}
\label{eq_sp_Gamma}
\left[{\sf h}^2\Delta  + \pi^2 \varrho_0^2 \Gamma^2 R_f^4  \Bigl({\hat I}_b  +  \sigma_{ph}^2\,
(\hat{b}^+ \hat{b} + 1/2)  \Bigr)^2\right]|\Psi\rangle ~=~0\,, 
	\end{equation}
where $|\Psi\rangle \in  \boldsymbol{H}_1$. 
{ Symbol $\Delta$ means Laplace operator in the space $L_2(D)$. This operator  is well-defined for
the functions $\Psi(\boldsymbol{r}) \in D_\Delta \subset L_2(D)$ which are doubly differentiable for
     $\boldsymbol{r} \nolinebreak \in \nolinebreak D\backslash\partial D$ and are continuous for
		$\boldsymbol{r} \nolinebreak \in \nolinebreak \partial D$. 
		We assume also that  homogeneous Dirichlet conditions are fulfilled  on the boundary $\partial D$ for all functions $\Psi(\boldsymbol{r}) \in D_\Delta$.}
Here, as in other places, we assume that the operator $\Delta$ is defined  in  the space $\boldsymbol{H}_1$   as $\Delta \otimes {\hat I}_b $.  The same remark applies to the operators in the second term of Eq.(\ref{eq_sp_Gamma}).
			Solving this  spectral problem, we find the spectrum of the value $\Gamma$:
		\begin{equation}
	\label{spectr_Gamma}
		\Gamma_{[m],n} ~=~ \pm\,\frac{{\sf h} R_f\lambda_{[m]} }{~\tilde\mu_0 L\bigl[ 1 +  \sigma_{ph}^2(n + 1/2)\bigr]}\,, 
				 \quad n = 0,1,2,\dots, N\,,
	\end{equation}	
		where  numbers $\lambda_{[m]}$ define the (dimensionless) eigenvalues for standard  Dirichlet problem
		\begin{equation}
		\label{Dir}
		 \Delta \Psi_{[m]}  ~=~  -  (\lambda_{[m]}/L)^2\Psi_{[m]}\,, \qquad \Psi_{[m]} \in  L_2(D)\,,\qquad \Psi_{[m]}(\boldsymbol{r})\Big\vert_{\boldsymbol{r}\in \partial D} = 0\,.
\end{equation}

{ As was noted above, the boundary set $\partial D$  is  piecewise-smooth surface.}
  As usual, we assume that the functions $\Psi_{[m]}(\boldsymbol{r})$ should be normalized to one.
It is clear that  the vectors $|\Psi_{[m],n}\rangle =  	|\Psi_{[m]}\rangle|n\rangle$ are the
eigenvectors of the spectral problem (\ref{eq_sp_Gamma}).

The paper \cite{Tal24_2} explores the asymptotic behavior of large numbers $[m]$, specifically for a certain type of domain.  For example, the case
\begin{equation}
\label{conv_circ}
\Gamma_{[m]} ~\longrightarrow~ \frac{{\sf h}\, {m}}{\mu_1} ~+~ {\mathcal O}({\sf h}^2) \,,
\end{equation}
is possible (the constant $\mu_1$ has the dimension of mass). 
Thus, the formula (\ref{conv_circ}) is asymptotically the same as the conventional formula $\Gamma_{m} = ({{\sf h}}/{\mu_1}){m}$, but it includes anomalous terms that are proportional to the constants ${\sf h}^n$, where $n = 2,3, \dots$.
Please note that the contradictions in the traditional formula for quantized circulation were discussed many years ago in the work  \cite{Donn} (see section 2.3.1 of this book).
We will not go into detail here. This topic has been thoroughly discussed in the author's previous works.

To understand the value of the number $N$ in equation (\ref{spectrum_R}), we need to define two auxiliary numbers, $N_1$ and $N_2$. These two numbers are determined based on the following criteria.
\begin{itemize}
\item[$N_1:$]  Approximate equality  $R_f\sqrt{1 + \sigma_{ph}^2\left(N_1 + \frac{1}{2}\right)} \simeq R_D$ holds, where $R_D$ will be radius of a maximal sphere inscribed in the considered domain $D$. In some cases, we can assume that $L = 2R_D$. Thus, the quantum number $n$ satisfies the uneqality:
\[ \sigma_{ph}^2 n ~<~ \left(\frac{R_D}{R_f}\right)^2\,;\]
\item[$N_2:$] Equality $ 1 +  \sigma_{ph}^2(N_2 + 1/2) = \sigma_{ph}^2N_2  + {\cal O}(1)$ holds so that the summand ${\cal O}(1)$ can be omitted. In this case the circulation  $\Gamma$   becomes  classical quantity because it stops being dependent on the  Planck's constant ${\sf h}$.
Therefore, if $N = N_2$ and $n > N$ in Eq.(\ref{spectrum_R}), the vortex ring becomes more like a classical object rather than a quantum one.	
Considering the previous point, we can say that the circulation shows quasi-classical behaviour when the quantum number $n$ meets the following condition:
\begin{equation}
\label{n_quasi}
1 ~<~ \sigma_{ph}^2 n ~<~ \left(\frac{R_D}{R_f}\right)^2\,.
\end{equation}
\end{itemize}

	Our next step is to find the energy spectrum of the dynamical system under consideration.
		Here we need to remember that energy is a physical value that is associated with time translations.
		Therefore, we must return to studying the evolution of the vortex loop in real time, $t$, rather than in ''conditional time'' $t^\#$:
		\begin{equation}
	\label{t_real}
		 t^\# ~\to~  t ~=~ \frac{4\pi R^2}{ t_0 |\Gamma|}\,t^\#\,.
		\end{equation}

		First, we consider the $t^\#$ - evolution.
		Quantized Hamiltonian $\widehat{H}^\#$ has following form:
		\begin{equation}
	\label{H_quant}
		\widehat{H}^\#  ~=~ -  \frac{{\sf h}^2}{2\mu_0}\Delta ~+~ \frac{{\sf h}\, \omega}{t_0}
		\left(b^+ b + \frac{1}{2}\right)\,.
		\end{equation}

		Spectral problem
		~$\widehat{H}^\# |\Psi\rangle ~=~ E^\#|\Psi\rangle$~
		has following solutions:
	\[ \boldsymbol{H}_1 \ni   |\Psi\rangle ~\equiv~  |\Psi_{[m],n}\rangle ~=~ 
	|\Psi_{[m]}\rangle|n\rangle\,, \qquad\quad
		|n\rangle = \bigl( \hat{b}^+ \bigr)^{n}|0\rangle \in \boldsymbol{H}_b\,,\] 
		where  the vector $|\Psi_{[m]}\rangle \in  L_2(D)$  is the eigenvector of the spectral problem
		(\ref{Dir}).
				Taking into account formula (\ref{eq_sp_Gamma}),
		the eigenvalues $E^\#_{[m],n}$ are written as follows: 
		\begin{equation}
	\label{E_eigen1}
	E^\#_{[m],n} ~=~  \frac{{\sf h}^2 \lambda^2_{[m]}}{2\mu_0 L^2} ~+~
	\frac{{\sf h}\,\omega}{t_0}\, \left(n + \frac{1}{2}\right)\,,
	\end{equation}
	where $[m]$ is multi-index and $n = 0,1,\dots,N$.
{ Generally, operator  $\widehat{H}^\#$    may have degenerate eigenvalues. For example, let the numbers $[m],n$ and $[\ell],k$ be fixed. In this case, equality $E^\#_{[m],n}  =  E^\#_{[\ell],k}$  can be seen as certain constraint  for the central charge  $\mu_0$  and constant $\omega$. 
 In exceptional cases,  parameters $\mu_0$  and  $\omega$  can satisfy this relationship  for individual vortices.  Accordingly, any vector
$|\Psi\rangle  =  A_1|\Psi_{[m]}\rangle|n\rangle  ~+~  A_2 |\Psi_{[\ell]}\rangle|k\rangle $
will be eigenvector    for arbitrary complex numbers  $A_1$ and  $A_2$. However, we will consider this case as an exception. Indeed,  as  the next step, we will study  a system of the $K$ vortices with arbitrary and different  constants $\omega = \omega^k$. In general, the numbers $\omega^k$ are the random numbers.  Moreover, the central charge $\mu_0$ is a fundamental constant in our model, and it will be determined based on experimental data in the future.  }

Let us consider the $t^\#$ - evolution of any vector $|\Psi\rangle \in {\boldsymbol H}$:
		\begin{equation}
	\label{t_ev1}
		\exp\left(\frac{i\widehat{H}^\# t^\#}{{\sf h}} \right)|\Psi\rangle ~=~  
		\sum_{[m],n}C_{[m],n}\exp\left(\frac{iE^\#_{[m],n} t^\#}{{\sf h}} \right)|\Psi_{[m],n}\rangle\,,
		\end{equation}
	where  $ C_{[m],n} = \langle\Psi_{[m],n}|\Psi\rangle$.	
				Next, we  restore the real time $t$ in this formula using Eq.(\ref{t_real}).
				We assume that following equality takes place for the real energy $E_{[m],n}$: 
				\[  E^\#_{[m],n} t^\#  ~=~  E_{[m],n} t\,.\]
				Therefore, 				the ''real - time'' evolution of any vector $|\Psi\rangle \in {\boldsymbol H}$   is written as
		\[ |\Psi\rangle ~\longrightarrow~ |\Psi(t)\rangle ~=~
		\sum_{[m],n}C_{[m],n}\exp\left(\frac{iE_{[m],n}\,t}{{\sf h}} \right)|\Psi_{[m],n}\rangle\,, \]	
					where 
		\begin{eqnarray}
	\label{energ_1}
 E_{[m],n} &=& \frac{{\sf h}^3 \lambda^3_{[m]}}{8 \pi \mu_0 \tilde\mu_0 v_0 L^2 R_f \bigl[1 +  \sigma_{ph}^2(n + 1/2)  \bigr]^2} ~+~\nonumber\\[2mm]
~&+& \frac{{\sf h}^2\,\lambda_{[m]}\omega\,  (n + 1/2)}{4\pi  \tilde\mu_0  R_f L\bigl[1 +  \sigma_{ph}^2(n + 1/2)  \bigr]^2 }\,
\end{eqnarray}

Let's write this formula in a different way. For this purpose, we introduce the following notation
for quantum energy unit in our model:
\[  E_{{\sf h}}   ~=~   \frac{{\sf h}^2}{4\pi\tilde\mu_0 R_f L }\,, \]
which depend both fluid property and size of the domain $D$.
We can use the values of the constants $\varrho_0$, $R_f$ and $L$ to calculate the $E_{{\sf h}}$ approximately. Let's assume that these constants have the following values: $\varrho_0 \approx 10^2\, Kg/m^3$ ({we used liquid $He_4$ density as a reference point}); $R_f \approx 10^{-9}\, m$ (atom size); $L \approx 10^{-2}\,m$. Under such assumptions, we find the value $E_{{\sf h}} \approx 10^{-32}\,J$. 

In these notations, formula (\ref{energ_1}) would be written as follows:
\begin{equation}
	\label{energ_2}
 E_{[m],n} ~=~ E_{{\sf h}}\,\frac{\lambda_{[m]}}{\bigl[1 +  \sigma_{ph}^2(n + 1/2)  \bigr]^2}
\left[\omega\, \biggr(n + \frac{1}{2}\biggl) ~+~ \frac{\sigma_{ph}^2}{2}\lambda_{[m]}^2  \right]\,.
\end{equation}

The second term in square brackets can be seen as a "fine structure" of energy levels because it is proportional to the value of ${\sf h}$.  Therefore, the second term in square brackets
  becomes significant only for very large values of the eigenvalues $\lambda_{[m]}$.
So, in our model, the Hamiltonian of a quantum vortex loop takes the following form:
\[ {\widehat H}(\omega) ~=~ \sum_{[m],n}E_{[m],n}|\Psi_{[m],n}\rangle\langle\Psi_{[m],n}|\,.   \]

\section{Physical analysis}

To make our analysis more visual, we will consider the domain $D$ to be a cube with an edge of length $L$. In this case
$[m] = (m_1, m_2, m_3)$ and
\[ \lambda_{[m]} ~=~ \pi\sqrt{m_1^2 + m_2^2 + m_3^2}\,, \qquad m_1,\,m_2,\, m_3 = 1,2,\dots\]
Let's study at the asymptotic expressions for the specified cubic domain in formula (\ref{energ_2}).
\begin{enumerate}
\item[I.] Numbers $[m]$ are small,  number $n$ is satisfied to condition $\sigma_{ph}^2 n < 1$. In this case energy levels will be as follows:
\begin{equation}
	\label{asympt_1}
 E_{[m],n} ~\longrightarrow~ E_{{\sf h}}\,\frac{\,\pi\,\omega\, \bigl(n + {1}/{2}\bigr)\sqrt{m_1^2 + m_2^2 + m_3^2}\,}{\bigl[1 +  \sigma_{ph}^2(n + 1/2)  \bigr]^2}
\, ~+~ {\cal O}({\sf h}^3) \,,
\end{equation}
where $n = 0,1,2,\dots,$ and $m_1,\,m_2,\, m_3 = 1,2,\dots, M$.  According to Eq.(\ref{asympt_1}), the distance between the nearby levels is as follows:
\[ E_{[m],n+1} - E_{[m],n} ~=~  E_{{\sf h}}\,\frac{\,\pi\,\omega\,\sqrt{m_1^2 + m_2^2 + m_3^2}\,}{\bigl[1 +  \sigma_{ph}^2(n + 1/2)  \bigr]^2} ~+~ {\cal O}({\sf h}^3)\,.\]
In this case, we have a hierarchy of levels that are not equally distant. As the number $n$ gets larger, the distance between levels decreases.
These levels are shown in Figure \ref{E_lev1}, where the constant $E_{{\sf h}}$ is used as the unit for energy and the constant $\omega  = 100$ is selected.  We will discuss the structure of the individual columns in Figure \ref{E_lev1} later.


\begin{figure}[H]
\begin{center}
		 {\hspace{40mm}~\includegraphics[width=6.5in]{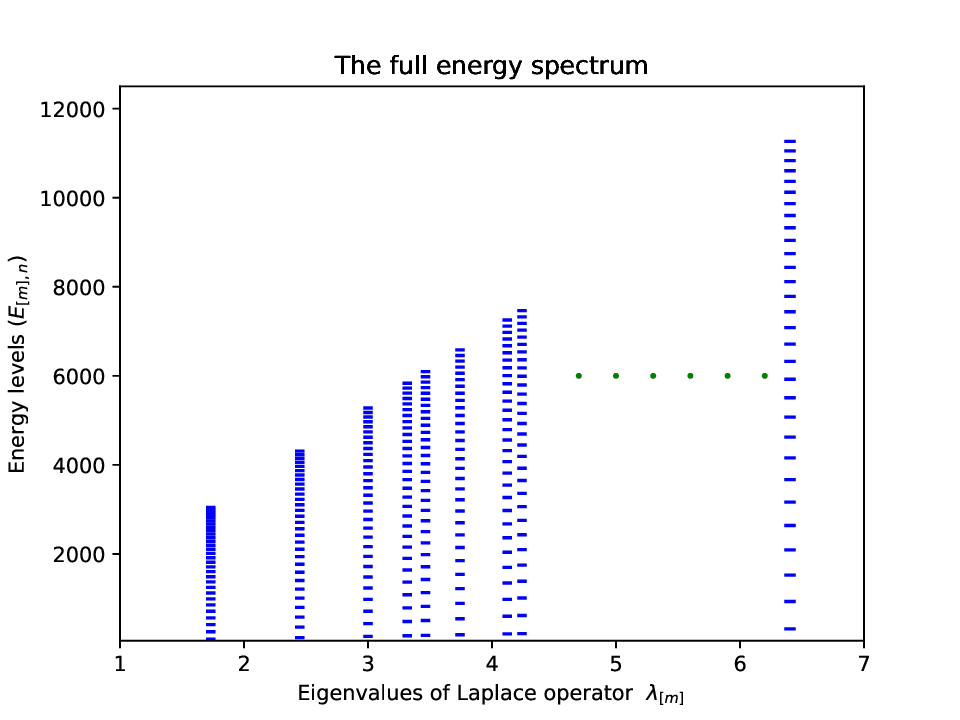}}
{\vspace{75mm}\captionof{figure}{Energy levels for the limiting case $I$.
\label{E_lev1}}}
\end{center}
\end{figure}

\item[II.] Numbers $[m]$ are small, number $n$ satisfies the condition (\ref{n_quasi}).
In this case we have following asymptotic behavior of the values $E_{[m],n}$:
\begin{equation}
	\label{asympt_2}
 E_{[m],n} ~\longrightarrow~ \frac{\,\mu_0^2 v_0^2 L\,\omega \sqrt{m_1^2 + m_2^2 + m_3^2}\,}{4 \tilde\mu_0 R_f (n + 1/2)}
 ~+~ {\cal O}({\sf h})\,.
\end{equation}
In this case, a term that does not depend on the Planck constant $h$ appears in the asymptotic expression for the energy.
Therefore,  the energy levels behave classically with only slight quantum corrections;
\item[III.] Numbers $[m]$ are large, number $n$ is small. In this case oscillatory modes will be suppressed: 
\begin{equation}
	\label{asympt_3}
	E_{[m],n} ~\longrightarrow~ \frac{\sigma_{ph}^2}{2} E_{{\sf h}}\,\pi^3 (m_1^2 + m_2^2 + m_3^2)^{3/2}\,.
\end{equation}	
In this case, the quantum effects will be the most noticeable because the condition
 $E_{[m],n} \propto  {\sf h}^3$ is holds;
\item[IV.] Numbers $[m]$ are large, number $n$ satisfies the condition (\ref{n_quasi}). In this case
\begin{equation}
	\label{asympt_4}
	E_{[m],n} ~\longrightarrow~ E_{{\sf h}}\,\frac{\,\pi^3 (m_1^2 + m_2^2 + m_3^2)^{3/2}\,}{2 \sigma_{ph}^2 n^2}
	\,.
\end{equation}	
In this limit, we observe a relation: $E_{[m],n} \propto {\sf h}$.
\end{enumerate}

Therefore, we can see that the quantum properties of the spectrum are manifested differently in different domains of the index set $\{[m], n\}$.

Let us consider a vortex in an arbitrary quantum state
\begin{equation}
\label{Psi_gen}
 |\Psi_C\rangle ~=~  \sum_{[m],n} C_{[m],n}|\Psi_{[m],n}\rangle\,,\qquad \sum_{[m],n} |C_{[m],n}|^2 = 1   \,.
\end{equation}
In this case
\[ E_\Psi ~=~   \langle\Psi_{[m],n}| {\widehat H} |\Psi_{[m],n}\rangle ~=~
\sum_{[m],n} |C_{[m],n}|^2 E_{[m],n}\,.\]
Considering the asymptotic behavior of the eigenvalues $E_{[m],n}$, we see that the dependence $E_\Psi = E_\Psi({\sf h})$ is not trivial. Moreover the function $E_\Psi({\sf h})$ depend on the quantum state $|\Psi\rangle$.

\begin{figure}[H]
		 {\hspace{10mm}~\includegraphics[width=6.5in]{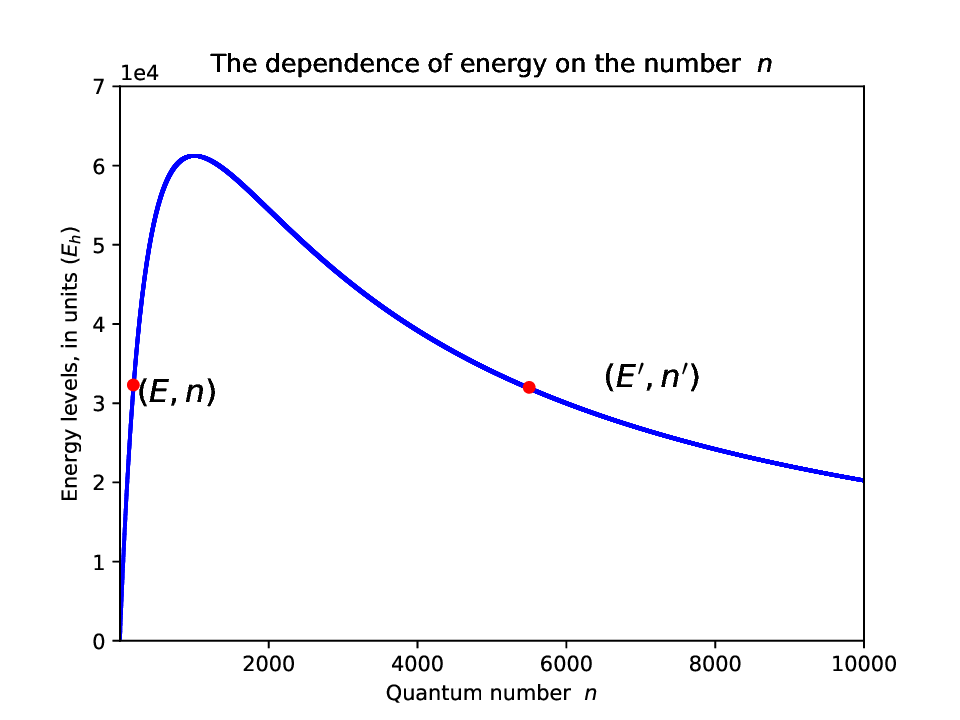}}
{\vspace{75mm}\captionof{figure}{The dependence of energy on the oscillatory number $n$.
\label{E_lev2}}}
\end{figure}

It is very interesting to investigate the energy spectrum of our system in the case where the eigenvalue $\lambda_{[m]}$ is fixed. This spectrum corresponds to  some  column in Figure \ref{E_lev1}.
For instance, let's consider the leftmost column, which corresponds to the lowest level of the Laplace operator. In this case, $m_1 = m_2 = m_3 = 1$. The result is displays on 
Figure \ref{E_lev2}.

The maximum of $E^{max}$  the curve in Figure \ref{E_lev2} is reached when the number $n$ as follows:
\[ n_{max} ~=~ \left[\,\frac{1}{\sigma_{ph}^2} ~-~ \frac{1}{2} ~-~ \frac{\lambda_{[m]}\sigma_{ph}^2}{\omega}\,\right]\,,\]
where square brackets in the notation  $[x]$ mean the integer pert of the number $x$.
It is worth noting that the value of $n_{max}$ lies outside the range defined in Eq.(\ref{n_quasi}), where the energy spectrum exhibits quasi-classical behavior.

The spectrum is displayed on the Figure \ref{E_lev2} has the following specific feature. 
The spectral curve\footnote{This curve, of course, is made up of a discrete set of points.}
contains the points $(E,n)$ and $(E^\prime,n^\prime)$ such that
\begin{equation}
\label{two_points}
  E ~\approx~  E^\prime\,, \qquad   n ~<< ~ n^\prime\,.
	\end{equation}
In the case $E << E^{max}$ the point $(E^\prime,n^\prime)$ lies in the domain that was defined by Eq.(\ref{n_quasi}). { It is important to note that the exact equality $E = E^\prime$ can only occur in exceptional cases (as it was discussed in the remark after Eq.(\ref{E_eigen1})).}
Due to the Heisenberg uncertainty principle, the energy levels in the spectrum always have a finite width $\delta E$  . Considering this, we will assume that for these points in the spectrum, the equality $E = E^\prime$ holds { if  $|E - E^\prime|\le \delta E$.}  
The circulation $\Gamma$ depend on the number $n$ in accordance Eq.(\ref{spectr_Gamma}).
Therefore, our model suggests that there are quantum states with ''almost equal''  energy levels but significantly different circulation values.

\section{$K$-vortex system}

Now we are considering a system that consists of $K$ non-interacting micro-vortices in a cube with an edge of length $L$. These vortices have  a random  (space)  coordinates that are defined by the vectors ${\boldsymbol q}_i \in D$, $i = 1,2, \dots, K$. { Each vortex, identified by the number i, has its own internal state, which is determined by the  ''coordinate''  $\chi_i$.} 
The  space of the quantum states of this system is as follows:
\[\boldsymbol{H}_K ~=~  \underbrace{\,\boldsymbol{H}_1 \otimes\dots\otimes\boldsymbol{H}_1\,}_{K}\,.\]
The basis in this space can be  formed by the vectors
\begin{equation}
\label{K_basis}
 |\Psi_{< w >}\rangle ~=~  |\Psi_{[m^1],n^1}\rangle |\Psi_{[m^2],n^2}\rangle \dots |\Psi_{[m^K],n^K}\rangle\,,
\end{equation}
{ where symbol $<\!w\!>$ means the  multi-index
\[ <\!w\!>\,: \qquad \Bigl\{(m^1_1,\,m^1_2,\, m^1_3; ~n^1); ~~\dots\,,~~ (m^K_1,\,m^K_2,\, m^K_3; ~n^K) \Bigr\}\,.    \]
 These vectors  are eigenvectors of 
the full Hamiltonian which can be constructed as follows:}
\[   {\widehat H}_v ~=~ \sum_{k=1}^K 
 \underbrace{{\widehat I}\otimes{\widehat I}\dots\otimes{\widehat I}}_{k-1}\otimes\, {\widehat H}(\omega^k)\otimes \underbrace{{\widehat I}\otimes\dots\otimes{\widehat I}}_{k+1}\,, \qquad    (K+1 \equiv 0)\,.\]
The dimensionless ''frequencies'' $\omega^k$ are random numbers here.
Defining this operator, we take into account the random Hamiltonians method \cite{Goff}.
The energy levels of this system  are as follows:
\[ {\sf E}_{<w>} ~=~  \sum_{k=1}^K E_{[m^k],n^k}(\omega^k)\,. \]

{  Using the basis (\ref {K_basis}) , we can write the wave function of the system under consideration in the ($\boldsymbol {q},\chi$) - representation.  This function  has  a form that is characteristic of the many-body quantum theory:
\[ \Psi_{< w >} ~=~ \Psi_{< w >}(\boldsymbol{q}_1,\dots, \boldsymbol{q}_K; \chi_1,\dots,\chi_K)\,.\]
Any pure state of the considered system is a linear combination of these wave functions.}
As usual, we assume that the quantum state of a real system of $K$ micro-vortices is represented by the density matrix:
\[ \hat\rho ~=~ \sum_{< w >} \rho_{< w >} |\Psi_{< w >}\rangle \langle \Psi_{< w >}|\,,\]

Let this system have a certain energy $E$. 
Such a system is associated with a specific distribution ${\cal D}(E,\Gamma)$   of circulations:
\begin{equation}
\label{circ_set}
{\cal D}(E,\Gamma): \quad
\{\, \Gamma_{[m^1],n^1}  ,\, \Gamma_{[m^2],n^2}  ,\, \dots\,,\Gamma_{[m^K],n^K}  \,\}
\end{equation}
We do not consider the rearrangement of vortices here.
  Based on the above, we can conclude that for this value of $E$ in the set (\ref{circ_set}), there are $2^K$ combinations of ${\cal D}(E,\Gamma)$:     
\[ E ~\longleftrightarrow~  \{{\cal D}_1(E,\Gamma),\, {\cal D}_2(E,\Gamma),\, \dots,\,{\cal D}_J(E,\Gamma)\}\,, \qquad\quad  J = 2^K\,.\]
Therefore, the energy levels in this system are highly degenerate.

In the proposed approach, the circulation is the value that minimally considers the movement of the fluid around the vortex filament.  Therefore, when studying a turbulent motion, we will focus on the distribution of circulation rather than the velocity distribution of fluid particles in the flow.
This means that the random transition
\begin{equation}
\label{trans}
 {\cal D}_i(E,\Gamma) ~\longrightarrow~ {\cal D}_j(E,\Gamma)\,, \qquad  i,j \le J\,,
\end{equation}
that occurs in a random time moments models the turbulent flow in our approach.

The next question is: under what conditions are such random transitions possible? So far, we have been considering vortices that are randomly located in the domain D  and are not affected by any external forces. 
Of course, this situation is an idealization, even if we ignore the processes of reconnecting vortex filaments.  In this case 
\[ \langle\Psi_{[m],n^0}| \Psi_{[m],n^1}\rangle  ~=~ 0\,, \qquad  n^0 \not= n^1\,,\]
for any quantum numbers $n^0 \not= n^1$. This equality is conserved in dynamics because 
the time evolution only provide the phase multiplier  for the vectors $| \Psi_{[m],n}\rangle$. Therefore, there are no physical reasons for the random transitions (\ref{trans}) in this context.

The standard way to account for the effects of a large number of objects located randomly is by using the mean field approach (see, for example \cite{Lim}).
In our case, these objects are circular vortex loops. 
It is reasonable to assume that the mean field influences both the vortex coordinates ($\boldsymbol q$) and the vortex filament geometry, which in this case cannot typically maintain its circular shape.
Therefore, we must consider the Hilbert space
\begin{equation}
	\label{space_ext}
	\boldsymbol{H}_1  ~=~  \boldsymbol{H}_{pq} \otimes   \boldsymbol{H}_b \otimes   \boldsymbol{H}_j 
	\end{equation}
instead space (\ref{space_quant}).
The symbol $\boldsymbol{H}_j$ represents the Hilbert space of harmonic oscillators, which correspond to Kelvin waves on a circular loop \cite{Tal24_2}.
 In this context, we need to make the necessary replacement:
\[   {\widehat H} ~\longrightarrow~  {\widehat H} ~+~ \varepsilon\,{\widehat V}\,,\]
where operator ${\widehat V}$ defines mean field potential in some way.
In general, this operator must be defined in the space (\ref{space_ext}) but not only in the space (\ref{space_quant}).
The purpose of this article is not to provide detailed information about the $V$ operator.
Our only reasonable assumption is that the off-diagonal elements of this operator are not zero:
\[ \langle\Psi_{[m^0],n^0}|\, {\widehat V}\,|\Psi_{[m^1],n^1}\rangle  ~\not=~ 0\,, \qquad  n^0 \not= n^1\,.\]
This means that a quantum state with specific values for numbers $[m^k]$ and $n^k$ evolves in the following way:
\[ |\Psi_{[m^k],n^k}\rangle\Big\vert_{t=0}  ~\longrightarrow~ \sum_{i,j}c_{i,j}(t) |\Psi_{[m^i],n^j}\rangle\,, \]
where, in general,  $c_{i,j}(t)\not=0$ for $t>0$, $i\not=k$  and $j\not=k$.
Therefore, we expect random transitions (\ref{trans}) to occur for any small value of $\varepsilon$ because they conserve the system energy. 

\section{Concluding remarks}

In this paper, we explored the arbitrary configuration  of a system of micro-vortices that can form at the early stages of turbulent flow.
The main result of this work is the calculation and physical analysis of the energy levels in such a system within an arbitrary bounded  domain.
Our approach to determining the quantum energy spectrum of the system in question is as follows:

		
	\vspace{5mm}	{\Large
\[
\begin{array}{ccccc}
\framebox(55,20){ \scalebox{0.8}{{\sf LIA}}} &\hspace{-3mm}\xLongrightarrow[\text{~~~}]{\text{${\sf{t}\to \sf{t}^\#}$}} & \framebox(95,20){\scalebox{0.7}{\sf Dynamic Eq.(\ref{LIE_pert}) }} & {\not\Longrightarrow}  & ~~E_{[m],n}  \\[3mm]
 \Bigg\Uparrow 
\lefteqn{\begin{turn}{90}${\hspace{-8mm}{~\sf{a} \to 0}}$\end{turn}}
&& \Bigg\Downarrow\lefteqn{\begin{turn}{-90}{\hspace{-12mm}\scalebox{0.7}{ \sf hamilt. str.}}\end{turn} }\vspace{3mm}&&\Bigg\Uparrow
  \lefteqn{\begin{turn}{90}${\hspace{-8mm}{\sf{t}^\# \to \sf{t}}}$\end{turn}}\\[5mm]
	\framebox(75,20){ \scalebox{0.8}{{\sf Vortex ring}}} && \framebox(95,20){ \scalebox{0.8}{ \sf energy:\,$H ={\cal E}_{cl}$}} & \xLongrightarrow[\text{\sf quantization}]{\text{${\sf h}$}}\hspace{-4mm}  & ~~E_{[m],n}^\# 
    \end{array}
\]}
		
		\vspace{5mm}

	The structure of the energy spectrum, which is unique to our approach, explains why turbulent motion arises  even in a rarefied system of $K$ quantum  vortices.
	Furthermore, depending on the state of the space $H_K$, different vortices exhibit their quantum properties in varying degrees in such a system.
	
		Our goal was not to find out why this system of vortices forms in the flow.
	The author has already considered one of these possible scenarios in his paper \cite{Tal_PoF23}. This paper explains how a system of many vortices can develop from a single vortex that is created randomly.  In his future works, the author plans to return to this topic and also explore the various aspects of vortex interaction.

\end{document}